\documentclass{aa}
\usepackage{epsfig}
\usepackage{graphicx}
\usepackage{amsmath,amsfonts,amssymb}
\newcommand{\be}{\begin{equation}}
\newcommand{\ee}{\end{equation}}
\newcommand{\bea}{\begin{eqnarray}}
\newcommand{\eea}{\end{eqnarray}}

\begin{document}

\title{Kinetic power of quasars and statistical excess of
MOJAVE superluminal motions}

\subtitle{}

   \author{M. L\'opez-Corredoira\inst{1,2}, M. Perucho\inst{3}}

   \offprints{martinlc@iac.es}

\titlerunning{Kinetic power / MOJAVE}
\authorrunning{L\'opez-Corredoira \& Perucho}

\institute{
$^1$ Instituto de Astrof\'\i sica de Canarias,
E-38200 La Laguna, Tenerife, Spain\\
$^2$ Departamento de Astrof\'\i sica, Universidad de La Laguna,
E-38206 La Laguna, Tenerife, Spain\\
$^3$ Dept. d'Astronomia i Astrof\'\i sica, Universitat de Val\`encia, E-46100 Burjassot, Val\`encia, Spain }

\date{Received xxxx/ Accepted xxxx}


  \abstract
{}      
%
{The MOJAVE (MOnitoring of Jets in AGN with VLBA Experiments) survey contains 101 quasars with a total of 354 observed radio components that are different from the radio cores, among which 95\% move with apparent projected superluminal velocities with respect to the core, and 45\% have projected velocities larger than 10$c$ (with a maximum velocity 60$c$). We try to determine
whether this distribution is statistically probable, and we make an independent measure of the kinetic power required in the quasars to produce such powerful ejections.} 
%
{Doppler boosting effects are analyzed to determine the statistics of the superluminal motions.
We integrate over all possible values of the Lorentz factor the values of the kinetic energy corresponding to each component. The calculation of the mass in the ejection is carried out by assuming the minimum energy state, i.e., that the magnetic field and particle energy distributions are arranged in the most efficient way to produce the observed synchrotron emission. This kinetic energy is multiplied by the frequency at which the portions of the jet fluid identified as ``blobs'' are produced.
Hence, we estimate the average total power released by the quasars in the 
form of kinetic energy in the long term on pc-scales.}
%
{A selection effect in which both the core and the blobs of the quasar are affected
by huge Doppler-boosting enhancement increases the probability of finding
a jet ejected within 10 degrees of the line of sight $\gtrsim 40$ times above what one
would expect for a random distribution of ejection, 
which explains the ratios of the very high projected velocities given above.
The average total kinetic power of each MOJAVE quasar should be very high to obtain
this distribution: $\sim 7\times 10^{47}$ erg/s. This amount is much higher than previous estimates of kinetic power on kpc-scales based on the analysis of cavities in X-ray gas or radio lobes in samples of objects of much lower radio luminosity but similar black hole masses. 
The kinetic power is a significant portion of the Eddington luminosity, 
on the order of the bolometric luminosity, and
proportional on average to $L_{\rm rad}^{0.5}$, with $L_{\rm rad}$ standing for radio luminosity, although this correlation might be induced by Malmquist-like bias.} 
%
{}


\maketitle

\section{Introduction}
\label{.intro}

Apparent faster-than-light motion among different components of a quasar, called 
``superluminal motion'', has been detected for more than 40
years ago (e.g., Gubbay et al. 1969; Knight et al. 1971; Cohen et al. 1971; Whitney et al. 1971).
Its measurement has been enhanced with radio observations by the Very Long Baseline Interferometry (VLBI; e.g., Cohen \& Unwin 1984), and the latest surveys at the NRAO (National Radio
Astronomy Observatory) Very Long Baseline Array (VLBA) conducted to monitor the jet kinematics over 
several years (Lister et al. 2009a,b).
The standard interpretation of these phenomena is that these relativistic jets are observed with a small angle to the line of sight (Rees 1966) [see explanation in \S \ref{.basic}].
Narlikar \& Chitre (1984) pointed out that the probability of getting the necessary beaming of these events is low. However, a quantitative assesment of this probability should be based on large complete samples. This is one of the pending problems for quasi stellar objects (QSOs), as discussed by 
L\'opez-Corredoira (2011), to which we pay some attention in this paper.

Apart from analyzing the difficulties in understanding this superluminal motion, the 
calculation of the jet power is also crucial for the understanding of the physical processes taking place in active galaxies.
Extragalactic jets represent one of the most powerful events in the Universe, 
and there is little doubt about their relevant role in the 
evolution of the host galaxy and its environment (Fabian et al. 2006; McNamara et al. 2005; McNamara \& Nulsen 2007). They heat the interstellar and intergalactic media (ISM and IGM, respectively) via shocks (e.g., Zanni et al. 2005, Perucho et al. 2011) and/or mixing (McNamara \& Nulsen 2007; De Young 2010; and references therein). The amount of energy deposited 
depends directly on the jet power and the age of the source (e.g., Merloni \& Heinz 2007). 
Moreover, the jet power is thought to be related to the properties of the host galaxy, such as the mass of the supermassive black hole in its center and the amount of gas in its surroundings (e.g., Merloni \& Heinz 2007; Cattaneo et al. 2009).
There are, however, still some open problems and challenges regarding
these phenomena (see, for instance, the review by K\"onigl 2010).

There are several ways to estimate the kinetic power from observations of AGN jets from scales of parsecs to hundreds of kiloparsecs (see a review in Ghisellini 2011, \S 5). Kinetic power can be estimated by means of an analysis of the cavity in X-ray gas, which is assumed to be created by the jet (e.g., Young et al. 2002; Rafferty et al. 2006; Merloni \& Heinz 2007). It can also be derived from extended radio emission (e.g., Willott et al. 1999; Punsly 2005; Xu et al. 2009), in terms of the lobe energetics and estimated ages of the radio lobes (e.g., Rawlings \& Saunders 1991; Kino \& Kawakatu 2005; Ito et al. 2008), assuming that most of
the injected energy is converted into work done by the expanding radio source. The time scales used to obtain the kinetic jet power either from the X-ray observations or the extended radio emission on kiloparsec scales, is typically considered on the order of $\sim 10^7$ yr (Ma et al. 2008).
The results in radio are subject to large errors, such as the determination of the volume and age
of the lobes. The age measurement is based on the measured advance velocity of the hot spot, which is assumed to be constant even though numerical simulations have shown that this may 
not be the case (e.g., Perucho et al. 2011), or on the spectral aging (Punsly 2005), which has been claimed to possibly be inaccurate (Katz-Stone \& Rudnick 1997).
In all cases, these methods are indirect estimations of the kinetic power, which can be severely
underestimated if the kinetic power is dissipated in heating or doing other work than the
creation of cavities in the gas. 

Estimations of the kinetic power from the jet kinematics have been derived from radio VLBI
observations, on the scales of parsecs, and timescales of a few years
(e.g., Celotti et al. 1997; Ma et al. 2008; Gu et al. 2009). Celotti et al. (1997) and Gu et al. (2009) analyzed 
X-ray observations assuming synchrotron self-Compton radiation to be responsible for the X-ray emission. This provided a poor estimation of the kinetic power because of the variability of the sources and the different
epochs at which radio and X-ray fluxes were measured (Celotti et al. 1997). 
Ma et al. (2008) used radio observations to derive the gas density in the jet. 
Their kinetic power estimations correspond to the energy flux crossing 
a given section per unit time at the moment in which that jet is observed, and 
there may be variations in the density of the injected material in the jet (Perucho et al. 2008).

In this paper, we propose something similar but with a totally new method, and we apply it to a recent survey,
in which we wish to measure the average total
power released by the quasars in the form of kinetic energy in the long term but on pc-scales. 
Although radio-components represent
a fitting artifact without any intrinsic physical meaning, they are generally related to shock waves traveling through the underlying jet flow. 
In this work, we calculate some properties of these components treating them as physically independent entities, which represents a first-order
approximation. In our estimates of the energetics of the radio components, we do not include the internal energy, only the kinetic energy.
We derive the mass of the particles from the flux in radio (similar to Ma et al. 2008). 
The key point of our analysis is that we will directly calculate the kinetic
energy by estimating the mass of gas included in each radio component and, hence its kinetic energy rather
than a measurement of the energy flux crossing 
a given section per unit time, and we also include a rough estimate of the frequency of production of these 
``blobs'' in the observations.
Our analysis takes advantage of a recent
survey of superluminal sources, and we focus only on quasars, 
in contrast to the BL Lac analysis of Ma et al. (2008).
Hence, with this method, we provide a direct calculation of the kinetic energy from observational data
rather than analyzing the hypothetical ways in which this energy is dissipated (the creation of cavities in the gas or other features).

Section \ref{.data} of our paper describes the data that enables us to perform our statistical
analysis. The method of analysis is described in \S \ref{.method}, and the results of its application to the data are given in \S \ref{.results}. Finally, our discussion and conclusions are provided in \S \ref{.concl}.

\section{Data}
\label{.data}

Lister et al. (2009a,b) monitored at the frequency of 15 GHz all radio-loud active galactic nuclei (AGNs) whose total flux density in 15 GHz is over 1.5 Jy for declination $\delta \ge 0$, and over 2.0 Jy for $-20^\circ \le \delta <0$, excluding the zone with Galactic latitude $|b|<2.5^\circ $. In total, they observed 135 objects to produce the MOJAVE (MOnitoring of Jets in AGN with VLBA Experiments) survey. Each object was observed for a median of 15 epochs over a period of 13 years (1994-2007). The 15 GHz images have higher than one milliarcsecond resolution, corresponding typically to parsec-scales.

Doppler boosting produces a strong bias in the range of observed angles of the jets,
so we do see neither all AGNs nor all jets. Nonetheless, there is a rough completeness of
objects up to a given total flux density.
There may be some missing objects for the adopted flux density limit, but we assume that the observed objects (135) are more or less complete. Waldram et al. (2010) gives an
independent measurement of the counts in 15 GHz in an area of 520 deg$^2$, $\frac{dN}{dF}\approx 51(F/{\rm Jy})^{-2.15}$ Jy$^{-1}$sr$^{-1}$ (valid for a flux $F$ between 5.5 mJy and 1 Jy) or, by integrating
\begin{equation}
N\approx 44(F/{\rm Jy})^{-1.15}\ {\rm sr}^{-1}
\label{cuentas}
.\end{equation}
If we extrapolate this law to higher values of $F$, we find that we should observe around 207 sources in our area (6.01 sr in the northern cap up to 1.5 Jy; and 2.09 sr in the southern cap up to 2.0 Jy), which is somewhat higher than the present number of 135, but taking into account the uncertainties, that we perform an extrapolation, and that some missing sources are expected in (high extinction) Galactic plane regions, we may assume that completeness is more or less reasonable.

Among the 135 AGNs, a total of 101 objects are quasars with $0.15<z<3.40$ {(roughly half of the sample with $z<1$ and half of the sample with $z>1$). The classifications of these AGNs were done by Lister et al. (2009a), based on the classification of the optical counterparts, except for four objects that were not classified because they had no optical counterparts. This means that a quasar has broad emission lines in its optical spectrum. This subsample of 101 quasars are the data used in the analysis of this paper. Each object has several components: a main core (containing most of the flux) and other minor components, at least one of which moves with respect to the main core, and has a flux of over 5 mJy. All these motions are relativistic with projected radial linear velocities between 0.2$c$ and 59.1$c$. In total, the 101 quasars have 101 main cores and 354 ejected blobs, 335 of them are superluminal (projected velocity $>c$), and 158 of them have projected velocities larger than 10$c$.

\section{Relativistic beaming and its energetic requirements}
\label{.method}

\subsection{Basic kinematic equations}
\label{.basic}

Given a blob expanding from its core with a velocity $v$, its Lorentz factor is
\begin{equation}
\Gamma =\frac{1}{\sqrt{1-\beta ^2}}
\label{lorentz}
,\end{equation} 
with \[
\beta=\frac{v}{c}.\]
The value of $\Gamma $ is much larger than one for ultra-relativistic beams. As pointed out by Rees (1966), a superluminal motion of a jet aligned in a direction very close to the line of sight of the source core may be inferred to have an ''apparent superluminal''
projected linear velocity (in units of $c$: $\beta _{\rm app}=\frac{v_{\rm app}}{c}$), since
\begin{equation}
\beta _{\rm app}=\frac{\beta \sin \theta }{1-\beta \cos \theta }
\label{beta}
,\end{equation}
which reaches a maximum value of $\beta _{\rm app}=\sqrt{\Gamma ^2-1}$ [for $\theta =\sin ^{-1}(1/\Gamma )$], a number without limit, provided that $\Gamma $ is also unlimited. Figure \ref{Fig:betaapp} illustrates the way
in which equation (\ref{beta}) behaves. We assume here that the observed apparent velocity of the blobs, $\beta _{\rm app}$, corresponds to the physical motion of the gas in the radio component (Lister et al. 2009b), rather than that of a propagating shock wave. This provides an upper limit to the velocity of the flow, if those radio components are interpreted as shock waves. 

\begin{figure}
\vspace{1cm}
{\par\centering \resizebox*{6.5cm}{6.5cm}{\includegraphics{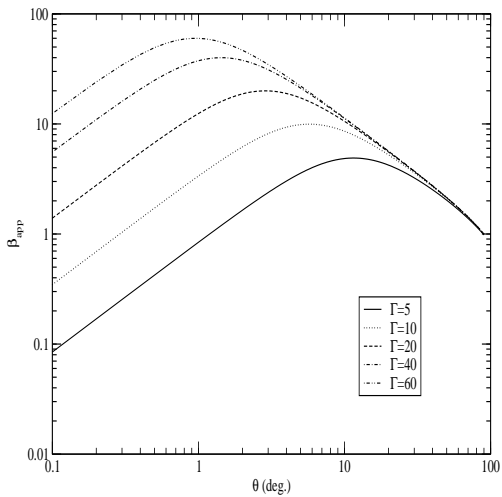}}
\par\centering}
\caption{Values of $\beta _{\rm app}$ for different values of $\theta $ and $\Gamma $
following Eq. (\protect{\ref{beta}}).}
\label{Fig:betaapp}
\end{figure}

Another property of the relativistic beaming is that the flux of the blob
is enhanced by Doppler boosting. The relationship between the observed flux ($F$) and the
intrinsic flux if the blob were at rest with respect to the quasar ($F_0$) is (Ryle \& Longair 1967; Narlikar \& Chitre 1984; Liu \& Zhang 2007)

\begin{equation}
F=\frac{F_0}{\left [\Gamma\left (1-\beta \cos \theta \right)\right]^{n_{\rm jet}-\alpha }}
\label{flux}
,\end{equation}
where $n_{\rm jet}=2$ or 3 depending on whether the jet is continuous or discrete, and 
$\alpha $ is the spectral index of the blobs with flux $F_\nu \propto \nu ^\alpha $.
This also explains why most times we only see the approaching jet and do not see the receding jet (counter-jet).
Fig. \ref{Fig:enhanc} illustrates the behavior of this equation for an approaching jet. As can be observed, Doppler factors smaller than one are also given for high values of $\theta$ in an approaching jet.

\begin{figure}
\vspace{1cm}
{\par\centering \resizebox*{6.5cm}{6.5cm}{\includegraphics{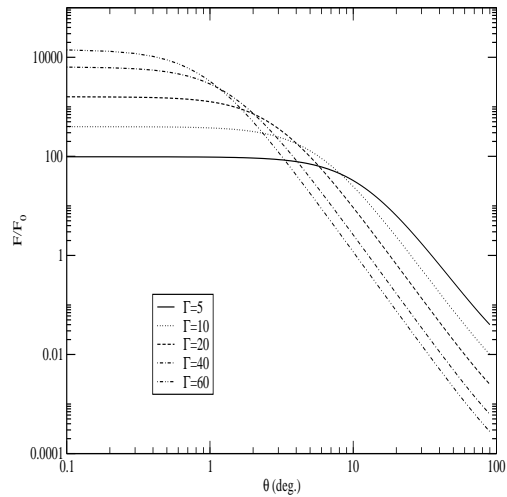}}
\par\centering}
\caption{Values of the enhancement of the flux in a blob (Doppler boosting), $F/F_0$, for different values of $\theta $ and $\Gamma $ following Eq. (\protect{\ref{flux}}), assuming $n_{\rm jet}-\alpha=2$.}
\label{Fig:enhanc}
\end{figure}

This relativistic beaming scenario clearly explains the data given by Lister et al. (2009a,b), and indeed Lister et al. (2009b) reproduce their observations using a model with some particular distributions of values of $\Gamma $, $\theta $, and $F_0$.
We may wonder whether this model is plausible from both a probabilistic and an energetic point of view.

\subsection{Probabilistic problem and selection effects}
\label{.selection}

When very few superluminal observed objects were known, one could be surprised to observe the low probability phenomena (Narlikar \& Chitre 1984). Now, observations are available for many sources up to a limiting flux density, and we see even more surprisingly that almost all the sources are superluminal, meaning that superluminity is not an anomaly/exception but the rule when observing the brightest blobs.

It is clear from Fig. \ref{Fig:betaapp} that, to obtain $\beta _{\rm app}>10$, as in almost half of the objects in our sample, one needs $\theta \lesssim 10^\circ $, and that this is nearly independent of the value of $\Gamma $ (for $\theta \gtrsim 10^\circ $, there are very small differences in the value of $\beta _{\rm app}$ between the case with $\Gamma =20$ and the case with $\Gamma =60$).
The probability of observing by chance an approaching jet with $\theta \lesssim 10^\circ $
is $\sim 0.015$; hence, randomly one would expect $5\pm 2$ blobs, out of 354, with
$\beta _{\rm app}>10$ for high values of $\Gamma $, 
but we observe 158, which is a totally improbable event by chance.
We need to explain an excess in a factor of $\sim 30$ in the number of sources with $\theta \lesssim 
10^\circ $ with respect to a random distribution. The number of 30 is a minimum ratio because
we assumed high values of $\Gamma $ and there may also be cases of  $\theta \lesssim 10^\circ $ for low $\Gamma $.
The blobs in each QSO are not independent since they have the same angle ejection, but in all cases
the statistics are valid: what we did is equivalent to taking a weighted distribution, with the weight given by the number of its blobs. We could do the statistics directly with the QSOs of average $\beta _{\rm app}>10$ in the blobs: randomly, one would expect $1.5\pm 1.2$ out of 101, and we observe 40, which is nearly the same ratio as before, but $\gtrsim 30$ times higher than in a random distribution. If instead the average $\beta _{\rm app}>10$, we required that the maximum $\beta _{\rm app}>10$ within a QSO, we would get 60 out of 101 QSOs, which is $\gtrsim 40$ times higher than in a random distribution.

We note that this estimation of the excess of probability is independent of the nature of the object, 
i.e. regardless of whether they are QSOs
or other kinds of AGNs. The classification of the same object may change if we see it with 
different orientations, but our estimation is merely based on the statistics of the probability of a lower
jet angle, independently of the kind of AGN.

The explanation stems from the selection effects in a magnitude limited sample: the Malmquist bias, by which we see systematically more luminous quasars at high redshift, but also an effect that leads to higher probabilities of observing low values of $\theta $, that is, accretion discs in the black holes that are nearly face on (Vermeulen \& Cohen 1994; Lister \& Marscher 1997). This last effect originates from Doppler boosting, the quasars being more luminous for low $\theta $ cases (see Fig. \ref{Fig:enhanc}). Since the limit of detection corresponds to obtaining a total flux (core+blobs) of 1.5/2.0 Jy (respectively, for positive and negative declination), in the cases with a core flux density lower than 1.5/2.0 Jy, we are biased towards a higher number of ejections with low $\theta$. 

The Doppler boosting affects both the observed blobs with some superluminal motion with respect to the core, and core itself. The light we see from the radio core partially originates from the jet.
The flux enhancement in the jet is a factor of around $10^2-10^4$ for $\Gamma $ between 5 and 60, and this has a strong effect on the probability distribution.
We assume an intrinsic distribution of quasar fluxes given by Eq. (\ref{cuentas}), which only approximately represents the intrinsic flux of the quasars because it is also affected by Doppler boosting. Nevertheless, it is adequate for a rough calculation. In addition, we assume that a fraction $l_{\rm jet}$ of the intrinsic light in the quasar (including core and blobs) originates from the jet [affected by the Doppler boosting of Eq. (\ref{flux})], whereas the remaining $1-l_{\rm jet}$ fraction stems from the part of the
quasar intrinsic emission  without relativistic motion. The probability distribution for observing an approaching jet with angle $\theta $ is then
\begin{equation}
\label{probtheta}
P(\theta )=A_P\sin \theta \left[1-l_{\rm jet}+\frac{l_{\rm jet}}
{\left [\Gamma\left (1-\beta \cos \theta \right)\right]^{n_{\rm jet}-\alpha }}\right]^{1.15}
,\end{equation}
where $A_P$ is a normalization constant such that the total probability for all angles is equal to one. In Fig. \ref{Fig:prob}, we plot the cumulative probability ($\int _0^\theta d\theta ' P(\theta ')$) for some values of $\Gamma $ and $l_{\rm jet}$. As can be observed, a cumulative probability up to 10 degrees of $\approx 0.5$ is obtained with several combinations of parameters.

\begin{figure}
\vspace{1cm}
{\par\centering \resizebox*{6.5cm}{6.5cm}{\includegraphics{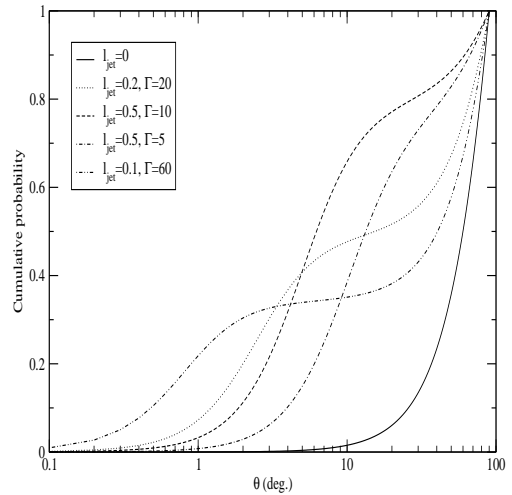}}
\par\centering}
\caption{Cumulative probability $\int _0^\theta d\theta ' P(\theta ')$, where $P(\theta )$ is from Eq. (\protect{\ref{probtheta}}).}
\label{Fig:prob}
\end{figure}

We note that to produce this important enhancement, the observed total density flux
should be dominated by light affected by a Doppler boosting: $\frac{l_{\rm jet}}
{\left [\Gamma\left (1-\beta \cos \theta \right)\right]^{n_{\rm jet}-\alpha }}>>1-l_{\rm jet}$.
This would not be possible if we assumed that the core light is unboosted, because in most
sources, most of the observed flux comes from the core. However, this assumption of
unboosted core light is incorrect because the radio flux 
we observe in the core originates mainly from the jet. Bell (2012) claimed that Doppler boosting cannot 
precisely explain these large ratios of superluminal motions because the core light
is unboosted, but we doubt the validity of Bell's statement.
Nevertheless, we refer the reader to the interesting discussion of Bell (2012).

\subsection{Estimate of the mass of the gas in the component}

The total mass embedded in the ejected blob can be calculated under the assumption that all its particles
produce synchrotron radiation (i.e. there is no thermal component) and the assumption of minimum energy, which states that the magnetic field and particle energy distributions are arranged in the most efficient way to produce the observed synchrotron emission. This approach thus provides a lower limit to the number of particles (and, therefore, of the mass) in the considered region. The radio luminosity of each blob is associated with its mass through (Perucho \& Mart\'\i \ 2002)

\begin{equation}
\label{masa_j}
M_j=\frac{L_{{\rm radio},j}C_1m _e}{C_3B_{\rm min}f_{\rm synch}}\frac{2+2\alpha }{2\alpha }
\frac{\nu _{\rm max}^\alpha -\nu _{\rm min}^\alpha }{\nu _{\rm max}^{(1+\alpha)} -\nu _{\rm min}^{(1+\alpha )}}
,\end{equation}    
\begin{equation}
\label{bmin}
B_{\rm min}=\left(\frac{6\pi AL_{{\rm radio},j}}{V_j}\right)^{2/7}
\end{equation}
\begin{equation}
\label{A_masa_j}
A=\frac{\sqrt{C_1}}{C_3}\frac{2+2\alpha }{1+2\alpha }
\frac{\nu _{\rm max}^{(1/2+\alpha )}-\nu _{\rm min}^{(1/2+\alpha )}}
{\nu _{\rm max}^{(1+\alpha)} -\nu _{\rm min}^{(1+\alpha )}}
,\end{equation}
where $C_1=6.3\times 10^{18}$ [c.g.s.], $C_3=2.4\times 10^{-3}$ [c.g.s.], $\nu _{\rm min}=10^7$ Hz, and
$\nu _{\rm max}=10^{11}$ Hz. The factor $f_{\rm synch}$ is the ratio
of the mass producing significant synchrotron emission, that is, the collective mass of the electrons or positrons (with mass $m_e=9.11\times 10^{-28}$ g), with respect to the total mass including also the protons. Protons do have a much lower emissivity, which can be assumed to be negligible.
$V_j$ is the physical volume of the blob, which we take as the volume of an ellipsoid
\begin{equation}
V_j=\frac{4}{3}\pi a_j^3r_j^2\times \left(\frac{\beta }{\beta _{\rm app}}\right)
,\end{equation}
where $a_j$ is the half width half maximum (FWHM/2) of the physical size (that is, multiplying the angular size by the angular cosmological distance), and $r_j$ is the axial ratio of the Gaussian ellipsoidal fit to the blob structure given by Lister et al. (2009b). The last factor
is the correction of the projection of the ellipsoid and its relativistic contraction, assuming that the jet is a moving bar/ellipsoid (Ghisellini 2000, \S 5.1). For the projected direction of the jet cross-section, we assume that the axis is equal to the one perpendicular to the direction of the jet propagation (i.e., we assume a cylindrical component). The volume used in the calculations is that obtained from the average value for all epochs.

The factor $\alpha $ is the spectral index, and $L_{{\rm radio},j}$ is the total radio
luminosity 
\begin{equation}
L_{{\rm radio},j}=4\pi d_L(z)^2(1+z)^{1+\alpha }F_{{\rm radio/rest},j}
,\end{equation}
where $d_L(z)$ is the luminosity distance of the quasar at redshift z (assuming a standard cosmology with $h=0.7$, $\Omega _m=0.3$, and $\Omega _\Lambda =0.7$) and $F_{{\rm radio/rest}, j}$ the total flux at rest with respect to the quasar (that is, corrected for Doppler boosting),
observed between $\nu _{\rm min}$ and $\nu _{\rm max}$. 
Using Eq. (\ref{flux}) and integrating the total density flux over the 
range of frequencies, we get

\begin{equation}
\label{fluxradio}
F_{{\rm radio/rest},j}=\frac {F_j}{\nu _0^\alpha (1+\alpha )}
[\nu _{\rm max}^{(1+\alpha)} -\nu _{\rm min}^{(1+\alpha )}]
\end{equation}\[ \times
\left[\Gamma _j\left(1-\beta _j\cos \theta _j\right)\right]^{n_{\rm jet}-\alpha }
,\]
where $F_j$ is the observed density flux at frequency $\nu_0=15$ GHz 
of the blob (we take the average of all epochs evaluated by Lister et al. 2009b), and
$\theta _j$ obtained from $\beta _{\rm app,j}$ and $\Gamma _j$ through
the relationship of Eq. (\ref{beta}):
\begin{equation}
\cos \theta _j=\frac{\frac{\beta _{{\rm app},j}}{\beta _j(\Gamma _j)}\pm\sqrt{1+\beta _{{\rm app},j}^2(1-\frac{1}{\beta _j(\Gamma _j)^2})}}
{1+\beta _{{\rm app},j}^2}
\label{theta}
.\end{equation}
There are two solutions to $\theta _j (\beta _{\rm app,j}, \Gamma _j)$, respectively, for the '+' and '-' sign in this equation. They correspond to the two values of $\theta $ for a given $\beta _{\rm app}$ and $\Gamma $ that are observed in Fig. \ref{Fig:betaapp}.  

In equations (\ref{masa_j}), (\ref{A_masa_j}), and (\ref{fluxradio}), there are values of $\alpha $ for which there is some zero in both the numerator and the denominator, but there is no divergence in the limit. For instance, for Eq. (\ref{masa_j}), $\lim _{\alpha \rightarrow 0} \frac{\nu _{\rm max}^\alpha -\nu _{\rm min}^\alpha }{\alpha }=ln \left(\frac{\nu _{\rm max}}{\nu _{\rm min}}\right)$.

\subsection{Kinetic energy of the observed blobs}

We do not know the value of $\Gamma $ for each blob, so we have to integrate over
all of its possible values. The kinetic energy of all the ``observed'' components in a quasar is (herein, we refer to as ``kinetic'' the total energy excluding the energy associated with the released mass at rest, that is $E=Mc^2(\Gamma -1)$)
\begin{equation}
\label{eq:ek}
E_{\rm K,obs.}=\langle Mc^2(\Gamma -1)\rangle
,\end{equation}
where
\begin{equation}
\langle f\rangle \equiv \sum _{j=1}^N\int _{\sqrt{1+\beta _{{\rm app},j}^2}}^{\Gamma _{\rm max}}d\Gamma _j
\end{equation}\[ \times 
\left[P_1(\Gamma _j|\beta _{{\rm app},j})
\sum _{\theta _j=+,-}P_2(\theta _j,\Gamma _j)f_j(\Gamma _j,\theta _j)\right]
,\]
and $j$ stands for the number of components from 1 to $N$, and $M_j$, $\Gamma _j$ are their respective masses at rest and Lorentz factors.
The probability of having a value of the Lorentz factor equal to $\Gamma _j$ provided that the apparent projected velocity in units of $c$ is $\beta _{\rm app,j}$, is (Bayes' theorem)

\begin{equation}
P_1(\Gamma _j|\beta _{\rm app,j})= C_P(\beta _{\rm app,j})\ P(\Gamma _j)P_{\rm beaming}(\beta _{\rm app,j}|\Gamma _j)
,\end{equation}
where $C_P$ is a constant of normalization of the probability over the range between
$\sqrt{1+\beta _{{\rm app},j}^2}$ (minimum value of $\Gamma $) and $\Gamma _{\rm max}$.
We take the Lorentz factor distribution from Liu \& Zhang (2007), which was derived precisely with MOJAVE data as well, with $P(\Gamma )\propto \Gamma ^{a_\Gamma }$ and $a_\Gamma =-1.73$.
The last factor stems from the probability of beaming in this range; that is, an amount 
proportional to the range of values $\theta _j$ that make $\beta _{\rm app}\ge \beta _{\rm app,observed}$. It is [Narlikar \& Chitre 1984, Eq. (8)]
\begin{equation}
P_{\rm beaming}(\beta _{{\rm app},j}|\Gamma _j)=\frac{1}{1+\beta _{\rm app,j}^2}\sqrt{\frac{\Gamma _j^2-1-\beta _{\rm app,j}^2}{\Gamma _j^2-1}}
.\end{equation}
In Fig. \ref{Fig:prob_gam}, we plot an example of the probability distribution for 
$\beta _{\rm app}=6.74$.

The sum ''$\sum _{\theta _j=+,-}$'' stands for the sum over the two possible values of $\theta (\Gamma _j, \beta _{{\rm app},j})$ given in Eq. (\ref{theta}) with the respective probabilities
\begin{equation}
P_2(\theta _j,\Gamma _j)=C_{P2}(\Gamma _j)\int _0^{\theta _j}d\theta \sin \theta 
\end{equation}\[ 
=C_{P2}(\Gamma _j)
(1-\cos \theta _j)
,\]
where $C_{P2}$ is a constant of normalization such that the sum of the two probabilities with sign $+$  and $-$ in Eq. (\ref{theta}) equals one.

\subsection{Total kinetic power}

To calculate the total power released, we must multiply each energy contribution by its frequency.
We note that the sample was observed over a period of 13 years. We assume statistically that the
average radial position during the period of the components ($R_j$) is in the middle of the observable
range, i.e., the range in which the jet might be observed to have sufficient flux density to be detected. We neglect that components get brighter or fainter during this period. While the sources in the sample are moving with an average angular speed $\mu _j$ (we neglect the acceleration), the order of magnitude of the frequency for each of the considered (or studied) events is
\begin{equation}
freq _j\sim \left(13+\frac{2R_j ({\rm \mu as})}{\mu _j({\rm \mu as/yr})}\right)^{-1}(1+z_j)\ {\rm yr^{-1}}
.\end{equation}
The $(1+z_j)$ factors stem from the time dilation caused by the cosmological expansion.
Including the frequency of events, the power released in the form of kinetic energy is the kinetic
energy of Eq. (\ref{eq:ek}) multiplied by the frequency
\begin{equation}
\label{kinpower}
p_{\rm K,total}\sim 2\left \langle freq\,M\,c^2\ (\Gamma -1)\right\rangle
.\end{equation}
The factor two accounts for the counter-jet. This counter-jet was not observed in any of the 101 quasars of our sample, which indicates that the viewing angles of all of them are small.

We cannot see the ejections for low value of $\Gamma _j$ because they have not undergone sufficient Doppler boosting to make them observable; to correct for this, one should divide this power by 
a factor of $P_3$ calculated as described in Appendix \ref{.lowenergy}. 
Furthermore, the jets are ejected in cones with very small intrinsic opening angles (Pushkarev et al. 2009 with MOJAVE data), so we can assume that
the correction for the unobserved blobs, required owing to the lower Doppler boosting in a cone, is negligible.
If the emission cones were wider, a correction should be introduced, as explained in 
Appendix \ref{.cone}. In this paper, we include the correction of the factors $1/P_3$ (derived in Appendix \ref{.lowenergy}) for each energy, but we neglect the factors $1/P_4$ (derived in Appendix \ref{.cone}).

\subsection{Parameters used for our calculations}

For the calculation of the kinetic energy/power, we need the parameters provided by Lister et al. (2009a,b) for each quasar and each component.

The value of $\alpha $, the spectral index, should be close to zero, since most of the
components have a flat radio spectrum (Lister \& Marscher 1997; Lister et al. 2009b). It is a rough
extrapolation to assume that this spectral index is valid for the range between $10^7$ and $10^{11}$ Hz, since we do not have information for the whole range and we know that these components are not optically thick in the whole range. This flat spectrum is most likely a combination of intrinsic emission and self-absorption. Nonetheless, as we show in \S \ref{.errors}, a different choice of $\alpha $ would not give significantly lower values
of the kinetic power.

The jet is considered continuous, so $n_{\rm jet}=2$ (Lister \& Marscher 1997; Lister et al. 2009b). 
The value of minimum flux density for the detection of a blob is, as we have said, $F_{\rm min}=5$ mJy.
For the maximum allowed value of the Lorentz factor, we take $\Gamma _{\rm max}=60$, given that the
maximum observed value of $\beta _{\rm app}$ is around that value.

A critical parameter is the fraction of emitting particles, 
which requires knowledge of the
composition of the plasma in the blob. This is composed of electron-positron pairs and 
electron-proton pairs. There have been several analyses indicating that the number
of electrons is $\lesssim 10$ times the number of protons in the jet (Kataoka et al. 2008;
Ghisellini \& Tavecchio 2010; and references therein).
A blob composed only of electron-positron pairs would decelerate strongly and
not allow superluminal motion to occur (Ghisellini \& Tavecchio 2010), owing to both a 
loss of energy in an inverse Compton process and the lower mass/unit charge that would
give a lower inertia to the jet.
Assuming the minimum possible number of protons (which contribute to a minimum kinetic energy), we take the number of 10 electrons per proton, hence the ratio of the mass of particles producing significant synchrotron emission to the total mass is $f _{\rm synch}=\frac{19m_e}{19m_e+m_{\rm proton}}=0.0103$.

\section{Results}
\label{.results}

We apply the aforementioned calculation to the quasar QSO 0016+731 (the first one in the list) 
at redshift $z=1.781$. It has only one observed jet with $\langle F_1\rangle =144$ mJy, $\langle R_1\rangle =1100$ 
$\mu $as, $\mu _1=87$ $\mu $as/yr, $\beta _{{\rm app},1}=6.74$, $a_1=330$ $\mu $as, and $r_1=1.00$. 
This gives a distribution of Lorentz factors similar to those given in Fig. \ref{Fig:prob_gam} and an average value of $\langle \Gamma _1 \rangle =20.8$, and the average angle with the line of sight $\langle \theta _1\rangle =15^\circ $. The average values within this distribution are $E_{\rm K,obs.}=4.33\times 10^{55}$ erg=24 M$_\odot c^2$, $\langle M_1\rangle=0.75$ M$_\odot $, volume of $V_1=77$ pc$^3$, and $E_{\rm K,total}=1.03\times 10^{56}$ erg. That is, the observed blob has a mass of around three quarters of a solar mass and a kinetic energy of around 24 M$_\odot c^2$, whereas the total released energy is 2.4 times higher including the jets that we do not see.
The frequency of an event such as this is estimated to be 
$freq _1\sim 0.073$ yr$^{-1}$ and consequently $p_{\rm K,total}\sim 2\times 10^{47}$ erg/s.

\begin{figure}
\vspace{1cm}
{\par\centering \resizebox*{6.5cm}{6.5cm}{\includegraphics{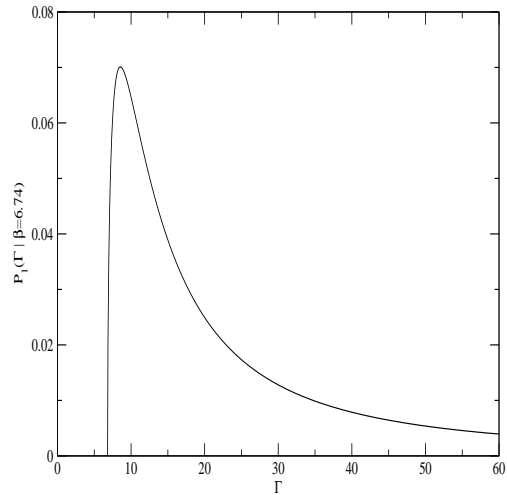}}
\par\centering}
\caption{Probability distribution of the Lorentz factor for a jet with an observed projected linear velocity (in units of $c$) of $\beta _{\rm app}=6.74$, as in the case of QSO 0016+731.}
\label{Fig:prob_gam}
\end{figure}

One might think that the QSO 0016+731 is a statistically anomalous case in which we have observed far more
superluminal jets than on average expected. To avoid this suspicion, we do the same calculation
for the rest of the 101 quasars. The histogram of Fig. \ref{Fig:histopk} shows the distribution of the minimum
values of $p_{\rm K,total}$. 
As can be observed, the case of QSO 0016+731 with $p_{\rm K,total}\sim 2\times 10^{47}$ erg/s is quite normal, and is quite close to the median value. 
There is large dispersion of values up to $p_{\rm K,total}\sim 3\times 10^{49}$ 
erg/s, which is reached for QSO 2037+511 at $z=1.686$ 
with one observed blob, and $\beta _{{\rm app},1}=3.3$. 
The average power and jet angle of the whole sample are
\begin{equation}
\overline {p_{K,total} } =(7.5\pm 2.9)\times 10^{47}\ {\rm erg/s}
,\end{equation}   
\begin{equation}
\overline{\theta }=(12.1\pm 1.4) \ {\rm deg}
.\end{equation} 

\begin{figure}
\vspace{1cm}
{\par\centering \resizebox*{6.5cm}{6.5cm}{\includegraphics{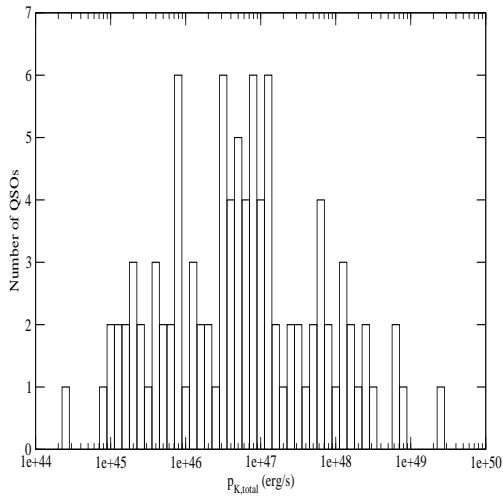}}
\par\centering}
\caption{Histogram of the distribution of kinetic power derived with Eq. 
(\protect{\ref{kinpower}}) for
the 101 quasars of the MOJAVE complete sample (Lister et al. 2009a,b).}
\label{Fig:histopk}
\end{figure}

\subsection{Correlation of kinetic power with the luminosity of the quasar}

In Fig. \ref{Fig:power_LQSO}, one can see the dependence of the kinetic power on the radio luminosity for the quasar $L_{\rm radio/rest,QSO}$, where we have considered the average 15 GHz flux of the core and we applied Eq. (\ref{fluxradio}) with $\alpha _{\rm QSO}=0$ but without the Doppler boosting correction ($\Gamma =1$, 
$\beta =0$). There is a significant correlation in this figure of $\frac{\langle xy\rangle}{\langle x\rangle \langle y\rangle}-1=(13.1\pm 3.7)\times 10^{-5}$
(significant at 3.5-$\sigma $), where $x\equiv \log_{10}L_{\rm radio/rest,QSO}$ (erg/s), and $y\equiv \log_{10}P_{K,total}$ (erg/s). However, there is also strong correlation both between $x$ and redshift
and between $y$ and redshift (the range is $0.15<z<3.40$), as expected due to the Malmquist bias, so this correlation mostly means that the most distant objects have the highest luminosities and highest kinetic powers. If we assumed that there is no evolution (i.e., there is no dependence of
quasar luminosities and kinetic powers on redshift), we could interpret the correlation
of Fig. \ref{Fig:power_LQSO} as something inherent to the characteristics of the quasars.
The best power-law fit $p_{\rm K,total}=K_R\ \left(\frac{L_{\rm radio/rest,QSO}}{10^{46}\ {\rm erg/s}}\right)^{\beta _R}$ gives $K_R=(5.0\pm 1.1)\times 10^{46}$ erg/s, and $\beta _R=0.47\pm 0.13$, which might be interpreted as the relationship under the assumption of non-evolution.

\begin{figure}
\vspace{1cm}
{\par\centering \resizebox*{6.5cm}{6.5cm}{\includegraphics{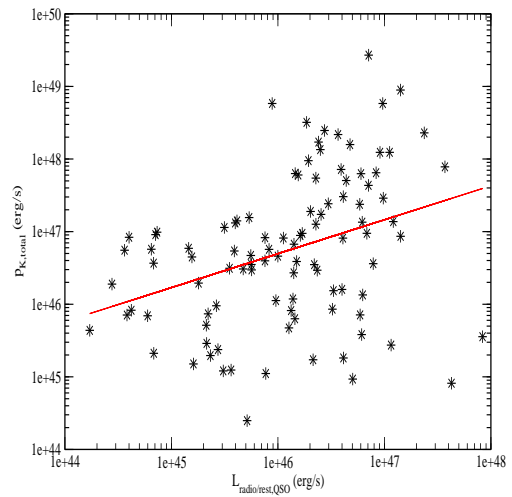}}
\par\centering}
\caption{Kinetic power released in jets versus the radio luminosity of the parent quasar.
The solid line is the fit indicated in the text.}
\label{Fig:power_LQSO}
\end{figure}

The dependence on the mass of the black hole can be evaluated in some cases in which the virial mass of the black hole is calculated. There are 24 QSOs in common with the sample of the Sloan Digital Sky Survey in the Data Release 7 (SDSS-DR7; Schneider et al. 2010), and in which the black hole virial masses were calculated (Shen et al. 2011). We take their 
fiducial values and compare with our kinetic power in the 24 QSOs in common in Fig.
\ref{Fig:power_MBH}. There is a slight correlation in this figure of
$\frac{\langle xy\rangle}{\langle x\rangle \langle y\rangle}-1=(3.3\pm 1.7)\times 10^{-4}$
(significant at 1.9-$\sigma $), where $x\equiv \log_{10}M_{\rm BH}(M_\odot)$, $y\equiv \log_{10}P_{K,total}$ (erg/s), which we consider not significant enough to take further conclusions.

\begin{figure}
\vspace{1cm}
{\par\centering \resizebox*{6.5cm}{6.5cm}{\includegraphics{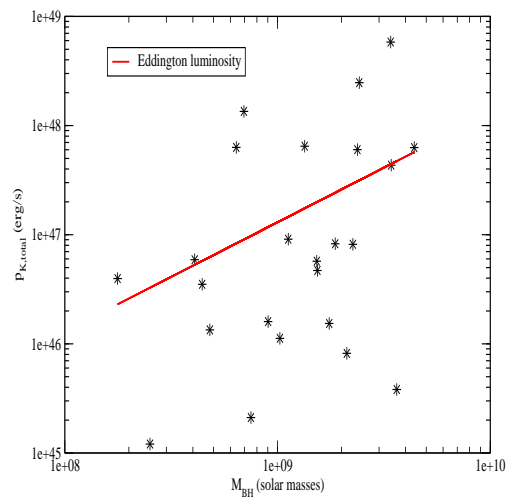}}
\par\centering}
\caption{Kinetic power released in jets versus the mass of the black hole of the parent quasar
for the cases in common between MOJAVE sample and SDSS-DR7. The solid line indicates the Eddington luminosity.}
\label{Fig:power_MBH}
\end{figure}

One way of determining the bolometric luminosity is to use the luminosity in the visible, and multiply it by a factor of the bolometric correction. We carried this out with R-band fluxes,  (obtained in NED database\footnote{http://ned.ipac.caltech.edu/}) or V or B when the first was unavailable, and we converted them into rest fluxes deriving the bolometric luminosity by means of the relation given by Runnoe et al. (2012)

\begin{equation}
L_{\rm bol}=0.75L_{\rm iso},
\end{equation}\[
L_{\rm iso}=\zeta (\lambda ) \lambda L_\lambda 
,\]\[
\zeta (\lambda )=4.25-3.63\times 10^{-4}\lambda (\AA )+2.27\times 10^{-7}\lambda (\AA )^2
,\]
where $\lambda $ is the wavelength at rest and $L_\lambda $ its corresponding luminosity per
unit wavelength.
The expression of an average $\zeta (\lambda )$ corresponds to a fit to the three values given by
Runnoe et al. (2012, \S 4.1). The result is shown in Fig. \ref{Fig:lumbol}. As can be observed,
the dispersion is high, most likely owing to the spread in the values of the real bolometric correction with respect to the given conversion.

\begin{figure}
\vspace{1cm}
{\par\centering \resizebox*{6.5cm}{6.5cm}{\includegraphics{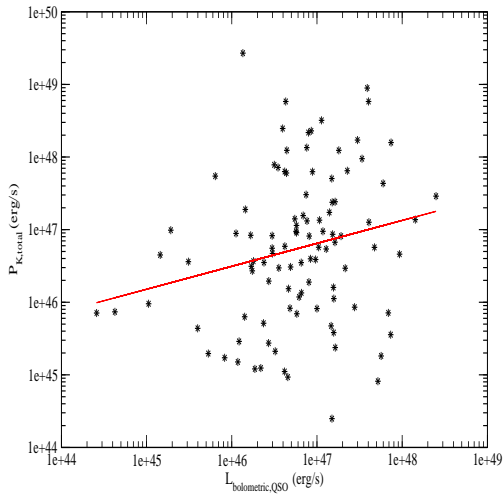}}
\par\centering}
\caption{Kinetic power released in jets versus the bolometric luminosity derived from
visible luminosities with bolometric corrections. The solid line is the fit indicated in the text.}
\label{Fig:lumbol}
\end{figure}

In figure \ref{Fig:lumbol}, the correlation is
$\frac{\langle xy\rangle}{\langle x\rangle \langle y\rangle}-1=(7.2\pm 3.3)\times 10^{-5}$
(significant at 2.2-$\sigma $), where $x\equiv \log_{10}L_{\rm bolometric,QSO}$ (erg/s), and $y\equiv \log_{10}P_{K,total}$ (erg/s). The correlation is barely significant (97.2\% C.L.), but let us accept that the correlation is there, and again assume that there is no evolution.
From a fit to Fig. \ref{Fig:lumbol}, we get $p_{\rm K,total}=(1.0\pm 0.2)\times 10^{47}\left(\frac{L_{\rm bol}}{10^{47}\ {\rm erg/s}}\right)^{0.32\pm 0.14}$ erg/s. 

The Eddington ratio is $\epsilon \equiv \frac{L_{\rm bol}}{L_{\rm Eddington}}$, 
where $L_{\rm bol}$
is the bolometric luminosity and the Eddington luminosity 
$L_{\rm Eddington}=1.3\times 10^{38} M_{\rm BH} (M_\odot )$ erg/s.
The dependence of $\epsilon $ on redshift is non-existent or negligible, but there is a 
dependence on luminosity for the QSOs for $z<5$ (L\'opez-Corredoira \& Guti\'errez 2012),
given by $\epsilon \approx 0.22\left(\frac{L_{\rm bol}}{10^{47}\ {\rm erg/s}}\right)^{0.35}$.
With this relationship and the above result, we get that the average kinetic power is 
around a fourth of the Eddington luminosity for a source with $L_{\rm bol}\sim 10^{47}\ 
{\rm erg/s}$, and the dependence on the luminosity is small 
($\frac{p_{\rm K,total}}{L_{\rm Eddington}}\propto L_{\rm bol}^{-0.33\pm 0.14}$).
We also note that some source may have super-Eddington kinetic power (see Fig. \ref{Fig:power_MBH}).
These are general results applicable to the bright radio-loud quasars in our present sample. 
 
Other correlations with other luminosities could also be explored, for instance, the correlation
with the luminosity of broad emission lines, which is approximately equal to the core
radio luminosity (Celotti et al. 1997).
Apart from the mass/luminosity and/or redshift, there is also a dependence of the kinetic power on the Bondi accretion rates (Allen et al. 2006), which we did not explore here.

\subsection{Possible sources of errors}
\label{.errors}

The calculation of the average kinetic power depends on the value of the different parameters and, although the preferred
values were chosen, we can explore how this average power changes with other parameters. In Table \ref{Tab:difpar}, we give the values of $\overline{ p_{K,total}}$ for different values of the parameters, except
for $f_{\rm synch}$: we show the dependence of the average 
total kinetic power on $n_{\rm jet}$, $\alpha $, $F_{\rm min}$, $\Gamma _{\rm max}$, and 
$a_\Gamma $.

\begin{table*}
\begin{center}
\caption[]{Values of the average kinetic power for different sets of parameters.
The standard values adopted in this paper correspond to the first row. Other parameters are used to test the sensitivity of the result to
the selected parameters; $f_{\rm synch}$ is kept to 0.0103.}
\label{Tab:difpar}
\begin{tabular}{ccccc|c}
\hline
$n_{\rm jet}$ & $\alpha $ & $F_{\rm min}$ (mJy) & $\Gamma _{\rm max}$ & $a_\Gamma $
& $\overline{p_{K,{\rm total}}}$ ($10^{47}$ erg/s) \\ \hline
\hline	    
2 & 0 & 5 & 60 & -1.73 & 7.5 \\
\hline
3 & 0 & 5 & 60 & -1.73  & 50.6  \\
2 & 1 & 5 & 60 & -1.73  & 2.5 \\
2 & 0.5 & 5 & 60 & -1.73  & 1.6 \\ 
2 & -0.5 & 5 & 60 & -1.73  & 80 \\ 
2 & -1.0 & 5 & 60 & -1.73  & 2000 \\
2 & 0 & 10 & 60 & -1.73  & 5.1 \\
2 & 0 & 1 & 60 & -1.73  & 2.3 \\
2 & 0 & 5 & $\max (\sqrt{1+\beta _{\rm app,j}^2}, 20)$ & -1.73  & 1.0 \\
2 & 0 & 5 & $\max (\sqrt{1+\beta _{\rm app,j}^2}, 40)$ & -1.73  & 3.6 \\
2 & 0 & 5 & 80 & -1.73  & 12.5 \\
2 & 0 & 5 & 100 & -1.73  & 18.7 \\
2 & 0 & 5 & 60 & -1.5 & 8.1 \\
2 & 0 & 5 & 60 & -2.0 & 7.0 \\
\hline
\end{tabular}
\end{center}
\end{table*}

The values of $f_{\rm synch}<0.0103$ (corresponding to fewer than 10 electrons per proton) can only increase the kinetic power since it is 
inversely proportional to $f_{\rm synch}$. Nonetheless, an underestimation of the ratio of electrons to protons (Kataoka et al. 2008; Ghisellini \& Tavecchio 2010) might also decrease the kinetic power. It is possible that the composition varies with the distance from the main core,
but in any case our values reflect an average expected composition on the observed scale. 

Another uncertainty might come from the calculation of the blob volume.
The beam size is $1.1\times 0.6$ mas, and the error in the position of the center
of the blob is 0.05 mas (Lister et al. 2009b). The error in the major axis $2a_j$ might
be a few tenths of mas with $2a_j\sim 1$ mas; that is, an error of a few tens per cent.
Since, from Eqs. (\ref{masa_j}), (\ref{bmin}), and (\ref{kinpower}), $p_{K,total} \propto a_j^{6/7}$, this error would not alter the order of
magnitude. In addition, since we only see the front part of the jet, there might a long tail behind that is less luminous due to a lower Doppler boosting; but this could only increase the volume, leading to a lower $B_{min}$ and higher mass of the jet; hence, a higher kinetic power.

We have certainly used here an assumption of the minimum energy, but a change in this hypothesis would not affect the order of magnitude either. All other methods of the kinetic power determination (see \S \ref{.intro}) also use either this assumption of minimum energy or the equipartition of energy.

We therefore find that a huge amount of kinetic power is obtained with any set of parameters. 
The most significant reduction (of an order of magnitude at most) could be obtained if we reduced the limiting $\Gamma _{max}$ to $\sim 20$ (but always over its minimum value $\sqrt{1+\beta _{\rm app,j}^2}$), or if we allowed as many as $\sim 100$ electrons per proton in the plasma.

\section{Comparison with other works, discussion, and conclusions}
\label{.concl}

Celotti et al. (1997), Ma et al. (2008), and Gu et al. (2009) obtain both a kinetic power of quasars on pc-scales on the same order that we obtained here, and similar correlations with the luminosity. Nonetheless, as said in \S \ref{.intro}, their kinetic power estimations correspond to energy fluxes (including both kinetic and internal energies) crossing the section of the observed jet per unit time at the moment in which it is observed, whereas our calculation tells us about the rate of total energy released in the form of kinetic energy by the quasar in the long term.

Rafferty et al. (2006) and Merloni \& Heinz (2007) compiled a list of measured kinetic powers on kpc-scales based on the nuclear properties of sub-Eddington AGNs, estimating the kinetic power from the $pdV$ work done to inflate the cavities and bubbles observed in the hot X-ray-emitting atmospheres of their galaxies and clusters.
Merloni \& Heinz (2007) obtained a dependence with the radio luminosity (uncorrected
of Doppler boosting, as in our case) similar to ours of $K_R=(8.7\pm 1.4)\times 10^{46}$ erg/s, 
and $\beta_R =0.54\pm 0.09$. However, the dependence on the black hole mass
is very different. If we take the data of kinetic power compiled 
in Table 1 of Merloni \& Heinz (2007), and we perform the fit versus black hole mass, 
we get $p_{\rm K,total}=(6.2\pm 3.6)\times 10^{43}\ 
\left(\frac{M_{\rm BH}}{10^9\ M_\odot}\right)^{1.62\pm 0.67}$ erg/s. Here, the exponent might be compatible, but the amplitude is much lower in Merloni \& Heinz than in our sample of Fig. \ref{Fig:power_MBH},
leading to a factor $10^3$ of difference.
This lower amplitude may be caused by the use of their sample of sub-Eddington AGNs,
which is a mixture of BL Lac objects and low Eddington ratio quasars, since our
quasars with higher average Eddington ratios are expected to have higher kinetic powers for the
same black hole mass.
However, it may also be caused by our direct measurement of the kinetic power giving
a much higher power than the direct measurement obtained by analyzing cavities in the X-ray gas.

Using radio observations, and assuming that radio lobes store most of
the kinetic power via work done by the expanding radio source, Xu et al. (2009) calculate the
kinetic powers of their sample of quasars. The average kinetic power for quasars with black hole masses of $10^9$ M$_\odot $ is  $\sim 5\times 10^{44}f^{3/2}$ erg/s, with $f$ a factor between 1 and 20 (Willott et al. 1999). Even for the highest allowed value of $f$, their kinetic power is lower than ours (see Fig. \ref{Fig:power_MBH}).
We said in Sect. \ref{.errors} that we could reduce an order of magnitude of value of the amplitude, for instance allowing 100 electrons/proton instead of 10.
Within an uncertainty of one order of magnitude for our kinetic power, our results fit the value of Xu et al. (2009). If our high values of the kinetic power on parsec-scales were confirmed,
it would imply that the extragalactic jets are more powerful than previously estimated from lobe energies and ages, or that there is a significant energy-loss from the parsec to the kiloparsec-scales. The latter is possible if the amount of energy in thermal particles in the lobes is underestimated by the equipartition assumption (e.g., Punsly 2005).
Nevertheless, we bear in mind that the different samples refer to different ranges
of redshifts and the evolution could also explain the differences.
In reality, the possibility that a large fraction of the energy in the lobes is in the form of thermal (invisible) particles has already been suggested (Ito et al. 2008) and evidence of metal-enriched gas has been found along the jets of several radio sources (Kirkpatrick et al. 2009, 2011). If the relation between these results is confirmed, we may be able to infer important implications for the mass-load of extragalactic jets. 

We have found that the kinetic power is a significant portion of the Eddington luminosity 
$\left[\frac{p_{K,total}}{L_{\rm Eddington}} \sim 0.2 \times 
\left(\frac{L_{\rm bol}}{10^{47}\ {\rm erg/s}}\right)^{-0.3}\right]$,
and on the order of the bolometric luminosity. 
It is proportional on average to $L_{\rm radio}^{0.5}$; the correlation
with bolometric luminosity is also similar, although less significant. In any case, these correlations might be affected by Malmquist bias or other selection effects. Once again, we recall that these results are valid for our sample of radio-loud quasars with relatively high luminosities and the range of redshifts between 0.15 and 3.4 (half of the sample with $z<1$ and half of the sample with $z>1$); we do not know whether these results can be extrapolated to other kinds of AGNs and whether some evolution might produce different results in different redshift ranges.

With regard to the probabilistic question mentioned in \S \ref{.intro}, we find from this analysis a distribution of superluminal sources observed by the MOJAVE collaboration (Lister et al. 2009b) in which from 354 observed blobs (down to 5 mJy in 101 quasars) 95\% are superluminal and 45\% are observed with projected velocities more than ten times the speed light.
There is a huge excess of jets ejected in a direction close to the line of sight. 
This can be explained as a selection effect in which both the core and the blobs are affected
by huge enhancements of fluxes produced by Doppler boosting, which makes the probability of finding
that a jet is ejected within 10 degrees of the line of sight $\gtrsim 40$ times higher than one would expect for a random distribution of ejections.

\

{\bf Acknowledgments:}
We are grateful for the helpful comments of Eduardo Ros,
Jos\'e Ma. Mart\'\i , and the anonymous referee.
We thank the text corrections by Claire Halliday (A\&A language editor).
This research has made use of data from the MOJAVE database 
that is maintained by the MOJAVE team (Lister et al. 2009a). 
This research has made use of the NASA/IPAC Extragalactic Database (NED),
which is operated by the Jet Propulsion Laboratory, California Institute of Technology, under contract with the National Aeronautics and Space Administration.
MLC was supported by the grant AYA2007-67625-CO2-01 of the Spanish Science Ministry.
MP acknowledges the financial support of the Spanish ``Ministerio de Ciencia e Innovaci\'on''
(MICINN) grants AYA2010-21322-C03-01, AYA2010-21097-C03-01, and CONSOLIDER2007-00050.

\

\appendix

\section{Total kinetic energy, including the unobserved ejections due to low energy}
\label{.lowenergy}

The energy $E_{\rm K,obs.}$ is the statistical average of the kinetic energy of the ejected blobs we have observed, but there are other ejections that cannot be observed because their flux density $F_j$ is lower than the detection limit $F_{\rm min}$. We can only see the ejections with high values of $\Gamma _j$ because the rest of them have insufficient Doppler boosting to make them observable.
The number of observed blobs correspond to the average number of expected blobs with $F_j>F_{\rm min }$,
where $F_{\rm min}$ is the minimum flux density required to be observable. 

Hence, the fraction of kinetic energy within the observed blobs (assuming 
$P(\Gamma )\propto \Gamma ^{a_\Gamma }$; Liu \& Zhang 2007) is
\begin{equation}
\label{P3}
P_3=\frac{\int _{\Gamma _{{\rm min},j}}^{\Gamma _{\rm max}}d\Gamma \ (\Gamma -1) P(\Gamma )}{\int _{1}^{\Gamma _{\rm max}}d\Gamma \ (\Gamma -1)P(\Gamma )}
=\frac{ \frac{\Gamma _{\rm max}^{a_\Gamma +2}-\Gamma _{{\rm min},j}^{a_\Gamma +2}}{a_\Gamma +2}-
\frac{\Gamma _{\rm max}^{a_\Gamma +1}-\Gamma _{{\rm min},j}^{a_\Gamma +1}}{a_\Gamma +1} }
{\frac{\Gamma _{\rm max}^{a_\Gamma +2}-1}{a_\Gamma +2}-\frac{\Gamma _{\rm max}^{a_\Gamma +1}-1}{a_\Gamma +1}}
.\end{equation} 
This minimum Lorentz factor $\Gamma _{{\rm min},j}$ at which the blob is observed is related to the minimum flux $F_{\rm min}$ and
the intrinsic flux density $F_{0,j}$ by means of Eq. (\ref{flux})
\begin{equation}
F_{\rm min}=\frac{F_{0,j}}{\left[\Gamma _{{\rm min},j}
\left(1-\beta (\Gamma _{{\rm min},j})\cos \theta _j\right)\right]^{n_{\rm jet}-\alpha }}
.\end{equation}
Using again Eqs. (\ref{flux}) for $F_{0,j}$, and Eq. (\ref{lorentz}), and solving
the quadratic equation (with the sign '-', which has the physical meaning of a minimum
$\Gamma $), we get
\begin{equation}
\Gamma _{{\rm min},j}=\frac{K_{\rm min} -\sqrt{\cos ^2\theta _j(K_{\rm min}^2-\sin ^2\theta _j)}}
{\sin ^2\theta _j}
,\end{equation}\[
K_{\rm min}=[\Gamma _j(1-\beta _j\cos \theta _j)]\left(\frac{F_j}{F_{\rm min}}\right)^
{\frac{1}{n_{\rm jet}-\alpha }}
,\]
provided that $\Gamma _{{\rm min},j}<\Gamma $, otherwise the values of
$\Gamma _j$, $\theta _j$ are not valid.

The total kinetic energy, both from the jets we see and the jets that we do not see, is the result of dividing the energy of each observed blob by $P_3(\Gamma _j,\theta _j)$. In the case of the data in this paper, for which the minimum flux density for the detection
of a blob is $F_{\rm min}=5$ mJy, we have calculated that $\overline{P_3}=0.92\pm 0.10$ with an r.m.s. of 1.0. We include this effect in our calculations in this paper because it is
important for some QSOs with low $P_3$.

\section{Conical jets}
\label{.cone}

If the jets were emitted in a cone with a significant width, we would have to
perform a new correction to the calculation of the total energy, since
the observed energy only accounts for the ejections with 
low values of $\theta _j$ and the rest of them have insufficient Doppler boosting to make them observable.
The number of observed jets correspond to the average number of expected jets with $F_j>F_{{\rm min },j}$,
where $F_{\rm min}$ is the minimum flux density required to be observable. 

\begin{figure}
\vspace{1cm}
{\par\centering \resizebox*{5cm}{11.9cm}{\includegraphics{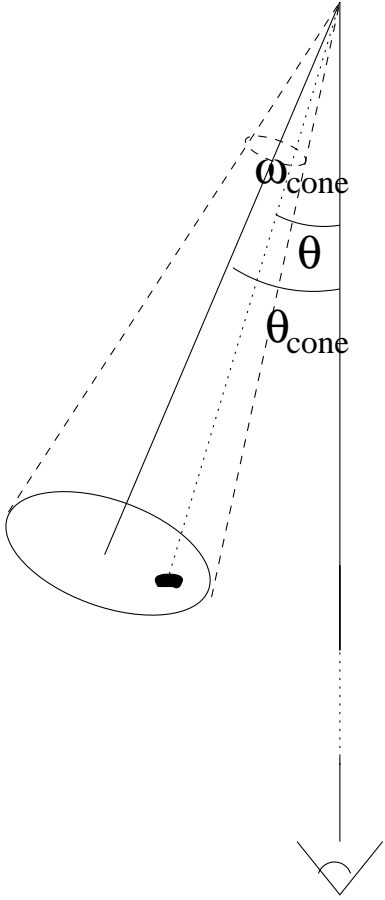}}
\par\centering}
\caption{Representation of the cone within which the jets move, and the blobs (filled black spot) are produced. The azimuthal angles ($\phi $, $\phi _{\rm cone}$) are not represented.}
\label{Fig:jet}
\end{figure}

The blobs are within a cone of solid angle $\omega _{\rm cone}$, as represented in Fig. \ref{Fig:jet}.
We must integrate over all possible values of the axis of the cone ($\theta _{\rm cone}$, $\phi _{\rm cone}$) for
a given approaching direction
($\theta _{\rm cone}<\pi/2$), and once the cone is fixed,
we integrate over all the possible directions of the ejection within that cone. Hence, the probability of observing an approaching jet is
\begin{equation}
\label{P4}
P_4(\Gamma _j, \theta _j)=\frac{1}{\omega _{\rm cone}^2}\int _{\Omega _1} 
d\Omega _{\rm cone} \int _{\Omega _2} d\Omega '
,\end{equation} 
where $\Omega _1\equiv $ conic solid angle around $(\theta _j, \phi _j=0)$ of $\omega _{\rm cone}$ stereo radians with $\theta _{\rm cone}<\pi/2$,\\
$\Omega _2\equiv $ conic solid angle around $(\theta _{\rm cone}, \phi _{\rm cone})$ of $\omega _{\rm cone}$ stereo radians,
such that $\theta <\theta _{{\rm min},j}$.

The minimum angle $\theta _{{\rm min},j}$ at which the blob is observed is related to the minimum flux $F_{\rm min}$ and the intrinsic flux density $F_{0,j}$ by means of Eq. (\ref{flux})
\begin{equation}
F_{\rm min}=\frac{F_{0,j}}{\left[\Gamma _j\left(1-\beta _j\cos \theta _{{\rm min},j}\right)\right]^{n_{\rm jet}-\alpha }}
.\end{equation}
Thus, using again Eqs. (\ref{flux}) for $F_{0,j}$, and Eq. (\ref{beta}), we get
\begin{equation}
\cos \theta _{{\rm min},j}=\frac{1}{\beta _j(\Gamma _j)}-\frac{\sin \theta _j}
{\beta _{\rm {\rm app},j}\left(\frac{F_{\rm min}}{F_j}\right)^{\frac{1}{n_{\rm jet}-\alpha }}}
.\end{equation}

Using spherical trigonometric relations, Eq. (\ref{P3}) can be rewritten as
\begin{equation}
P_4(\Gamma _j, \theta _j)=\frac{4}{\omega _{\rm cone}^2}\int_{\max [0,(\theta _j-\rho _{\rm core})]}^{\min[\pi /2, (\theta _j+\rho _{\rm core})]}d\theta _{\rm cone}\int _0^{\rho _{\rm cone}}d\theta _c'
\end{equation}\[
\times \sin \theta _{\rm cone}\sin \theta _c'
\]\[
\times \cos ^{-1}\left[\max \left(-1,\frac{\cos \rho _{\rm cone}-\cos \theta _{\rm cone}\cos \theta _j}{\sin \theta _{\rm cone}\sin \theta _j}\right)\right]
\]\[
\times \cos ^{-1}\left[\max \left(-1,\min \left(1,\frac{\cos \theta _{{\rm min},j}-\cos \theta _{\rm cone}\cos \theta _c'}{\sin \theta _{\rm cone}\sin \theta _c'}\right)\right)\right]
.\]
We note that $\theta _{\rm cone}$ is an angle with respect to line of sight, whereas $\theta _c'$ is an angle with respect to the axis of
the cone. The angular radius $\rho _{\rm cone}$ stands for the semiangle of the cone
\begin{equation}
\rho _{\rm cone}=\cos ^{-1}\left(1-\frac{\omega _{\rm cone}}{2\pi }\right) 
.\end{equation}

The kinetic power, both from the jets we see and the jets we do not see, results from dividing the energy of each observed blob by $P_4$. Nonetheless, given that from MOJAVE data $\rho _{\rm cone}=0.13/\Gamma <<1$ (Pushkarev et al. 2009), we find that $P_4\approx 1$ and the correction is negligible. Had $\rho _{\rm cone}$ a high value [for instance, in the model by Blandford \& K\"onigl (1979), $\rho _{\rm cone}=1/\Gamma $], we 
would have to apply this correction for low values of the Lorentz factor $\Gamma $.

\end{document}